\documentclass[trackchanges]{aastex701}

\usepackage{graphicx}
\usepackage[version=4]{mhchem}

\newcommand{\nautilus}{Nautilus}
\newcommand{\jwst}{\textit{JWST}}
\newcommand{\teq}{T_{\rm eq}}

\newcommand{\mearth}{M_\oplus}
\newcommand{\rearth}{R_\oplus}

\begin{document}

\title{Unveiling the Diversity and Origin of Sub-Neptunes with the Nautilus Space Observatory}

\author[orcid=0000-0003-0156-4564]{Luis Welbanks}
\affiliation{School of Earth and Space Exploration, Arizona State University, Tempe, AZ, USA}
\email[show]{luis.welbanks@asu.edu}

\author[orcid=0009-0006-1763-5936]{Kylie~E.~Hall}
\affiliation{Lunar and Planetary Laboratory, University of Arizona, 1629 E. University Boulevard, Tucson, AZ 85721, USA}
\email{kyliehall1@arizona.edu}

\author[orcid=0000-0003-2415-2191]{Julien de Wit}
\affiliation{Department of Earth, Atmospheric and Planetary Science, Massachusetts Institute of Technology, 77 Massachusetts Avenue, Cambridge, MA 02139, USA}
\email{jdewit@mit.edu}

\author[orcid=0000-0002-5322-2315]{Ana Glidden}
\affiliation{Department of Earth, Atmospheric and Planetary Science, Massachusetts Institute of Technology, 77 Massachusetts Avenue, Cambridge, MA 02139, USA}
\affiliation{Kavli Institute for Astrophysics and Space Research, Massachusetts Institute of Technology, Cambridge, MA 02139, USA}
\email{aglidden@mit.edu}

\author[orcid=0000-0003-3989-5545]{Noah Tuchow}
\affiliation{Lunar and Planetary Laboratory, University of Arizona, 1629 E. University Boulevard, Tucson, AZ 85721, USA}
\email{nwtuchow@arizona.edu}

\author[orcid=0000-0001-7962-1683]{Ilaria Pascucci}
\affiliation{Lunar and Planetary Laboratory, The University of Arizona, Tucson, AZ, USA}
\email[]{}  

\author[orcid=0000-0003-3714-5855]{D\'aniel Apai}
\affiliation{Steward Observatory, The University of Arizona, 933 N. Cherry Avenue, Tucson, AZ 85721, USA}
\affiliation{Lunar and Planetary Laboratory, University of Arizona, 1629 E. University Boulevard, Tucson, AZ 85721, USA}
\affiliation{Alien Earths Team, NASA ICAR/NExSS, USA}
\email{apai@arizona.edu}

\author[0000-0003-1127-8334]{Robin Wordsworth}
\affiliation{Harvard Paulson School of Engineering and Applied Sciences, 29 Oxford Street, Cambridge, MA 02138, USA}
\affiliation{Department of Earth and Planetary Sciences, Harvard University, 20 Oxford Street, Cambridge, MA 02138, USA}
\email[]{}

\author[orcid=0000-0002-3627-1676]{Benjamin V. Rackham}
\affiliation{Department of Earth, Atmospheric and Planetary Science, Massachusetts Institute of Technology, 77 Massachusetts Avenue, Cambridge, MA 02139, USA}
\affiliation{Kavli Institute for Astrophysics and Space Research, Massachusetts Institute of Technology, Cambridge, MA 02139, USA}
\email[]{brackham@mit.edu}

\author[orcid=0000-0001-5989-7594]{Chia-Lung Lin}
\affiliation{Steward Observatory, The University of Arizona, 933 N. Cherry Avenue, Tucson, AZ 85721, USA}
\email{chialunglin@arizona.edu}

\author[orcid=0000-0002-5887-1197]{Raymond Pierrehumbert}
\affiliation{Department of Earth, Atmospheric and Planetary Science, Massachusetts Institute of Technology, 77 Massachusetts Avenue, Cambridge, MA 02139, USA}
\email[]{}

\author[orcid=0000-0003-3204-8183]{Mercedes López-Morales}
\affiliation{Space Telescope Science Institute}
\email[]{email@stsci.edu} 

\author[orcid=0000-0000-0000-0001,sname='A.~D.~Feinstein']{Adina~D.~Feinstein}
\affiliation{Department of Physics and Astronomy, Michigan State University, East Lansing, MI 48824 USA}
\email[]{adina@msu.edu} 


\begin{abstract}

Sub-Neptunes are the most common class of planets in the Galaxy, yet they have no Solar System analog and remain poorly understood as a population. \jwst{} observations have revealed atmospheres spanning a wide range of metallicities, compositions, and cloud properties, driving active debates over whether warm sub-Neptunes harbor liquid water oceans beneath \ce{H2}-rich envelopes, maintain stratified \ce{H2}/\ce{H2O} interiors, or have well-mixed, metal-rich envelopes. Open questions also remain over what physical processes drive transitions between hazy and clear atmospheres. These are intrinsically population-level questions that single-target observations, however deep, cannot resolve. Here we argue that a sub-Neptune population survey with the Nautilus Space Observatory, a proposed constellation of large-diameter space telescopes, would deliver the first statistical map of sub-Neptune atmospheric diversity, test competing classification schemes, identify habitable candidates, and serve as a pathfinder population for the eventual habitable-worlds search. These goals are achievable across the proposed mission classes for the constellation, and this architecture is uniquely well-matched to this science case since population-level questions demand sample size and a uniform observing strategy.

\end{abstract}


\section{About Nautilus} 
\textit{This White Paper presents a potential science case for the Nautilus Space Observatory, a concept under development for a NASA Strategic Mission for the Astro 2030 Decadal Survey. Nautilus is a constellation of space telescopes and will provide a modular, scalable, sustainable, upgradable, expandable space observatory that can be deployed rapidly and then expanded progressively. The core concept for Nautilus is described in \cite{Apai2019}. This White Paper is part of the first series of science white papers capturing ideas that emerged from the Nautilus Science Case workshop (held at MIT in May 2026).} \newline

\section{Scientific Context and Problem Statement} 
Sub-Neptunes ($\sim 1.8$--$4\, \rearth$) are the most common class of planet in the Galaxy and outnumber super-Earths in the habitable zone of FGK stars when normalized over orbital period \citep{Bergsten2022}, yet have no Solar System analog and remain poorly understood as a population \citep{Bean2021}. Based on mass and radius measurements, their bulk densities are consistent with a continuum of interior structure compositions spanning rocky-cores, water-rich interiors, and \ce{H2}-rich envelopes \citep{Seager2007}.  In the James Webb Space Telescope era, transmission spectra of $\sim$10 sub-Neptunes have established that their atmospheres span a wide range of metallicities, mean molecular weights, and cloud properties \citep[e.g.,][]{Madhusudhan2023, Benneke2024, Davenport2025}. This emerging diversity is not yet matched by a coherent physical picture: competing classification schemes describe the population along different axes \citep{Madhusudhan2025PNAS, Benneke2024}, and the \emph{same} planet can remain consistent with multiple architectures even with the best available data. TOI-270\,d, with the highest signal-to-noise sub-Neptune spectrum to date, is interpretable as a hycean world \citep[i.e., a temperate planet with a liquid water ocean beneath a thin \ce{H2}-rich envelope; \citealt{Madhusudhan2021}; e.g.,][]{Holmberg2024}, a miscible-envelope sub-Neptune \citep{Benneke2024}, or a planet experiencing the interactions between a magma-ocean and an \ce{H2}-rich atmosphere \citep{Nixon2025}. K2-18\,b, the flagship hycean candidate, shows detected \ce{CH4} and \ce{CO2} but a conspicuous absence of \ce{NH3}, with interpretations spanning ocean dissolution, magma surface, photochemical destruction, and high-metallicity gas dwarf scenarios \citep[][]{Hu2025, Madhusudhan2025, Shorttle2024}.

A second, equally unresolved axis of diversity is the role of clouds and aerosols. Population studies find that the prevalence of high-altitude aerosols varies with equilibrium temperature \citep{Brande2024}. However, planets with otherwise similar bulk properties show qualitatively different atmospheric properties; temperate sub-Neptunes such as LP\,791-18\,c break the potential trend toward clearer atmospheres expected at lower equilibrium temperatures \citep{Roy2026}, suggesting that formation history and stochastic processes also imprint on the observable population. Resolving these questions (e.g., the architecture occurrence rates, the existence of critical transitions in chemistry and clouds, and the dependence of the population on host-star properties) requires a sample size far beyond what \jwst{} or any single-aperture facility can deliver, and \emph{uniform} characterization across that sample.

\textbf{Problem Statement:} Understanding the diversity and origin of sub-Neptunes is intrinsically population-level: it requires uniform transmission spectroscopy of tens to hundreds of sub-Neptunes across host-star type, mass, and equilibrium temperature, at a sample size and uniformity that current and planned facilities cannot provide. The Nautilus Space Observatory \citep{Apai2019}, a proposed scalable constellation of large-aperture space telescopes  (D$\approx$10--50\,m equivalent), is well-matched to delivering this survey across all of its proposed mission classes.


\section{Science Objectives} 

\begin{figure}[t!]
    \centering
    \includegraphics[width=1.0\textwidth]{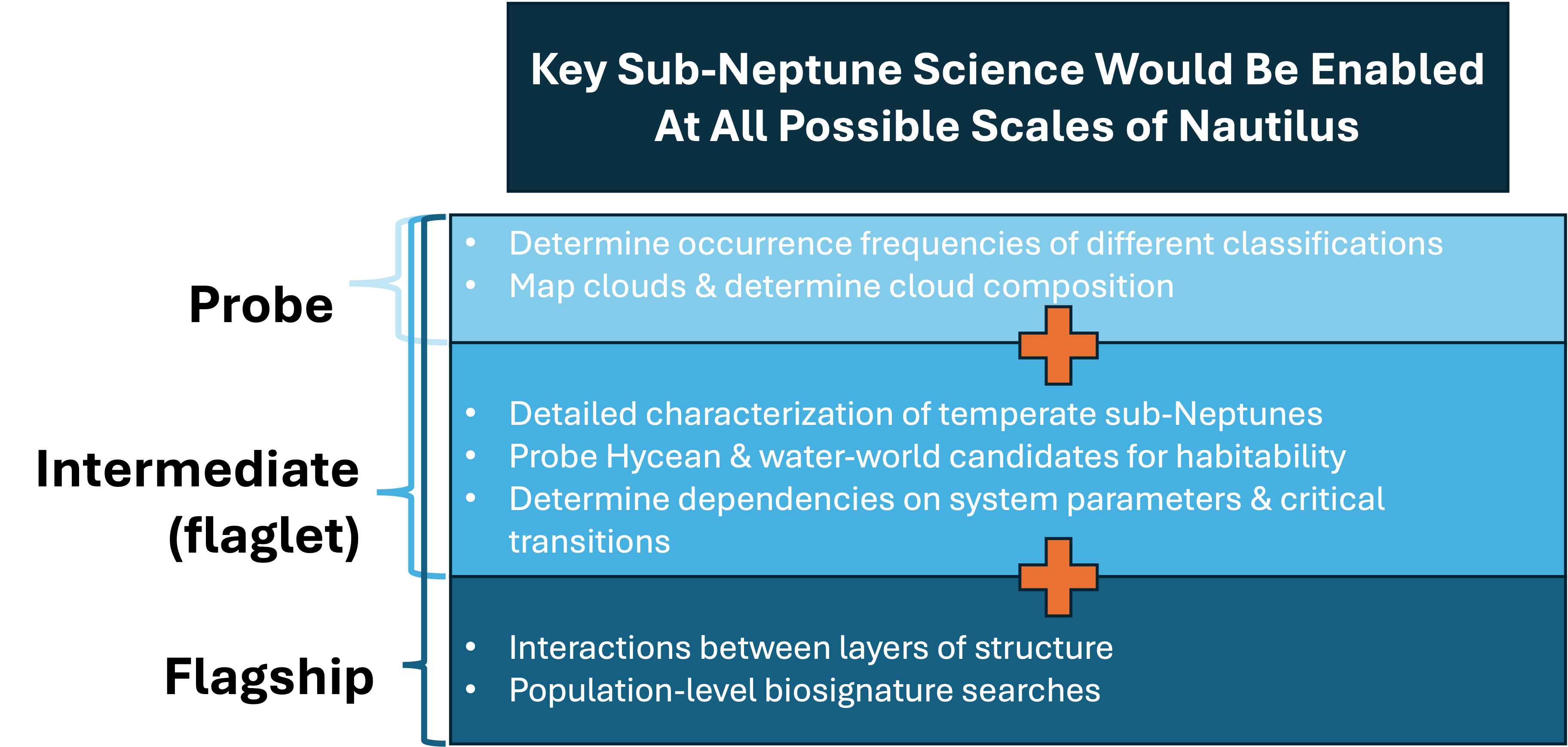}
    \caption{Sub-Neptune science deliverables across the proposed Nautilus mission classes. Each scale enables the science objectives of the smaller class plus additional capabilities, with the flagship configuration integrating sub-Neptune science into the broader habitable-worlds survey.}
    \label{fig:scales}
\end{figure}

The \nautilus{} sub-Neptune population survey would address the following objectives, with scientific reach scaling with mission class ( Figure \ref{fig:scales}).

\begin{enumerate}
    \item \textbf{Characterization:} Determine the atmospheric composition, mean molecular weight, and elemental inventory. Test atmospheric and interior architectures consistent with a liquid water surface, magma-ocean boundary, or supercritical interior. For instance, atmospheric signatures of ocean–atmosphere interaction \citep[e.g., CO\textsubscript{2} depletion relative to expectations for a dry envelope,][]{Triaud2024} provide one population-level diagnostic where accessible. Constrain the inferred bulk interior structure (\ce{H2O}, \ce{H2}, rock/metal mass fractions). 
      
    \item \textbf{Architecture occurrence rates:} Informed by the bulk and atmospheric properties above, Nautilus would determine whether the sub-Neptune population is described by a single archetype or by distinct sub-classes (e.g., gas dwarfs, miscible-envelope sub-Neptunes, steam worlds, etc), and measure the fraction of planets belonging to each class. These insights would help determine the classification scheme for sub-Neptunes.

    \item \textbf{Critical transitions:} Identify whether there is a critical boundary in irradiation/equilibrium temperature, envelope mass fraction, or other physical parameter, at which atmospheric chemistry or aerosol onset change character. The expected transitions include temperature-driven changes in dominant chemistry (e.g., the CH$_4$/CO transition near T$_{\rm eq}\sim1000$~K), stellar environment and its impact on photochemical reactions, as well as composition-driven transitions between hydrogen-dominated and heavier envelope architectures, including compositional changes across the radius valley \citep[e.g.,][]{Cherubim2025}.

    \item \textbf{Population trends:} Test whether the gas-giant mass--metallicity relation extends into the sub-Neptune regime, breaks, or shows a discontinuity; whether cloud/haze prevalence tracks equilibrium temperature, composition, host-star UV, or formation history; and whether architecture distributions depends on host spectral type, system architecture, orbital separation or other. 
    
    \item \textbf{The habitable subset:} Identify which sub-Neptunes are plausibly habitable (if any) and at what occurrence rate. Habitability indicators may include temperate equilibrium temperatures consistent with surface liquid water and chemistry consistent with a water-bearing surface or hydrosphere. These planets are the natural bridge between the sub-Neptune population science and the eventual habitable Earth-like exoplanet search.
    
\end{enumerate}

Delivering on these objectives also requires understanding the physical processes connecting atmospheres to formation and evolution including but not limited to atmospheric escape and photo-evaporation  \citep[which likely shape the radius valley separating super-Earths and sub-Neptunes;][]{Fulton2017}; the climate of warm and temperate worlds including \ce{H2O} condensation, internal heat flux, and the longitudinal and latitudinal distribution of clouds that shapes both albedo and the transmission spectrum \citep{Yang2013, Turbet2021}; magma--atmosphere interactions \citep{Kite2020}; and the host stars themselves (activity, UV environment, flares). 




\section{Data Requirements}

While initial reconnaissance of sub-Neptunes can be performed using broad-band photometry, delivering on the science objectives of the survey requires transmission spectroscopy of a large sub-Neptune sample. At the probe level, the wavelength coverage and spectral resolution can be modest to build an initial sample for the population. Detailed characterization at the intermediate (flaglet) or flagship level will require higher fidelity spectroscopy. The priorities for all mission classes are described below:
 
\begin{itemize}
\item \textbf{Wavelength coverage.} Continuous coverage from the optical through the near-infrared is required ($\sim 0.4$--$3\,\mu$m) as a baseline and $\sim 3$--$5\,\mu$m as a high-value extension. The baseline range probes Rayleigh and cloud/aerosol scattering signatures, alkali features, and the major molecular bands of \ce{H2O}, \ce{CH4}, \ce{CO2}, \ce{NH3}, and \ce{SO2}, supporting population-level classification, cloud/aerosol mapping, and major-molecule chemistry \citep[e.g.,][]{Benneke2013, Line2016, Welbanks2019}. The $3$--$5\,\mu$m extension enables access to the strongest \ce{CO2} and \ce{CO} bands which can be helpful in breaking degeneracies between composition and clouds. A dedicated retrieval and survey-design study is needed to quantify which wavelength ranges are required for each science objective and to distinguish the baseline requirements for population classification from enhanced capabilities for detailed abundance and degeneracy-breaking studies.

\item \textbf{Spectral resolution.} Moderate resolution ($R \sim 100$--$500$) is sufficient for the core population survey, identifying and characterizing strong molecular bands, retrieving abundance ratios and mean molecular weight, and identifying the He~I 1.083$\mu$m line as an escape diagnostic. Higher resolution ($R \gtrsim 30{,}000$) is an enhanced capability that would enable precise abundance measurements and overcoming degeneracies between overlapping molecular features (e.g., \ce{H2O} and \ce{CH4}). This approach may be particularly cost-effective if implemented in a design restricted to $\lesssim3\mu$m.

\item \textbf{Sensitivity.} Per-visit transit-depth precision sufficient to constrain order-unity scale-height features (typical scale heights of warm sub-Neptunes are 100--300\,km). For canonical targets around bright M and K dwarfs, this corresponds to $\sim$30--50\,ppm per spectral resolution element over a single transit. Lower-mass sub-Neptunes and fainter or more variable hosts require co-addition across multiple transits.
 
\item \textbf{Sample size and target selection.} A representative population survey requires uniform transmission spectroscopy of $\sim 50$--$100$ sub-Neptunes (probe class), scaling to several hundred (flaglet) and $\gtrsim 1000$ (flagship). Target selection should be informed by the range of properties along which competing sub-Neptune scenarios make different predictions, including equilibrium temperature ($\teq{} \sim 200$--$1000$\,K), planet mass ($\sim 2$--$15\,\mearth$), host spectral type (late M through G), UV / activity environment, orbital separation, and system architecture. Multi-transit observations co-added under a uniform pipeline are central to delivering the population-level precision required.
 
\item \textbf{Companion data.} Accompanying radial-velocity or TTV mass measurements to $\lesssim 20\%$ precision \citep[e.g.,][]{Batalha2019} are valuable for anchoring interior modeling and to convert transmission-spectroscopy constraints into interior architecture/composition inferences. However, for favorable high-SNR systems and instrument designs, the transmission spectra themselves may also provide useful constraints on surface gravity and planetary mass through the atmospheric scale height, especially when temperature, composition, and pressure structure are constrained jointly. The Nautilus survey should therefore treat mass information as coming from a combination of RV/TTV measurements, population priors, and TS-based mass constraints where applicable.

\item \textbf{Host-star characterization.} Stellar photospheric heterogeneity directly affects the reliability of transmission spectra, while chromospheric activity, UV irradiation, and flares shape atmospheric chemistry, escape, evolution, and aerosol production. The sub-Neptune survey should therefore be designed to test for correlations between stellar photospheric/chromospheric properties and planetary atmospheric properties, linking this program directly to the \nautilus{} stellar-heterogeneity and flare science cases.

\end{itemize}

\section{Analysis and Interpretation} 

The analysis proceeds in two coupled layers. At the \emph{per-planet} level, transit light curves obtained across multiple \nautilus{} units must be combined into a single, calibrated transmission spectrum, a non-trivial analysis challenge that requires careful treatment of inter-unit offsets, instrument systematics, and stellar contamination. Atmospheric retrievals based on parametric models \citep[e.g.,][]{Madhusudhan2018, Welbanks2021} and physically-informed self-consistent models \citep[e.g.,][]{Bell2023, Welbanks2024} then constrain molecular abundances, cloud-deck and haze properties, mean molecular weight, and (when paired with mass measurements) interior architecture \citep[e.g.,][]{Madhusudhan2020, Welbanks2024}. 
 
At the \emph{population} level, hierarchical Bayesian inference combines per-planet posteriors to constrain the joint distribution of architecture class, envelope metal mass fraction, cloud-top pressure, and host-star and orbital dependences. Success is defined by the recovery of population-level distributions at statistical power sufficient to distinguish among competing classification schemes (e.g., a $\geq 3\sigma$ preference for one partition of the population over another) and to identify trends and transitions in chemistry and clouds as a function of equilibrium temperature. Population-survey simulation tools \citep[e.g., Bioverse;][]{Bixel2021} provide a  framework for quantifying these statistical thresholds and for trading sample size against per-target observing depth.

\section{Relevant Science Requirements}

System-level science requirements are summarized in the Tables \ref{scireq1} and \ref{scireq2} below.

\section{Relevance to Nautilus and Mission Class} 

Sub-Neptunes are favorable targets for \nautilus{}: transit depths and atmospheric scale heights are large, and bright nearby M and K dwarf hosts are abundant in the TESS and forthcoming PLATO catalogs. Photon budgets per target are orders of magnitude more advantageous than for the Earth-twin case, and the population-level science requires large sample sizes that no single-aperture facility can deliver on a realistic timeline.

\textbf{Constellation.} A constellation directly addresses the two binding constraints. The large volume of uniform transit-spectroscopy observations can be distributed across multiple units operating in parallel, removing target accessibility as the bottleneck. A staged deployment also enables a survey-then-follow-up strategy, in which later or differently configured units revisit the most scientifically valuable targets identified by the initial survey.

\textbf{Relevant Class.} Sub-Neptune science scales naturally across the proposed mission classes (Fig.~\ref{fig:scales}):

\begin{itemize}
\item \textbf{Probe.} $\sim 50$--$100$ sub-Neptunes; first map of architecture occurrence and cloud distributions; identification of temperate water-world candidates.
\item \textbf{Flaglet.} Several hundred targets; statistical characterization of the habitable-candidate subset; quantitative tests of mass--metallicity and cloud/chemistry transitions.
\item \textbf{Flagship.} Sub-Neptune science extends as the pathfinder population for the thousand-Earth biosignature survey \nautilus{} was conceived for, derisking the architecture, operations, and population-inference methodology.
\end{itemize}


\section{Relevance to NASA and Astrophysics Strategy} 


Strategic Objective 1.2 of the 2022 NASA Strategic Plan is to ``understand the Sun, solar system, and universe," and aligns with NASA's emphasis on ``improving techniques and ideas for discovering and characterizing habitable and/or inhabited environments" on exoplanets. Sub-Neptunes are the predominant outcome of planet formation in the Galaxy, and understanding their diversity is fundamental to placing the Solar System, which conspicuously lacks a sub-Neptune, in galactic context.

This science case also directly advances the Astro2020 Decadal Survey priority ``Worlds and Suns in Context" \citep{Astro2020} and the questions E-Q2 (\emph{What are the properties of individual planets, and which processes lead to planetary diversity?}) and E-Q3 (\emph{How do habitable environments arise and evolve within the context of their planetary systems?}). Sub-Neptunes are a substantial component of the small-planet population from which Habitable Worlds Observatory (HWO) reflected-light targets will be drawn, and characterizing their transit-accessible counterparts complements rather than competes with HWO: the two facilities sample different orbital and host-star regimes and together produce the population-level framework that any biosignature search will require for interpretation.

\begin{acknowledgments}
We thank the Heising-Simons Foundation for supporting the Nautilus Science Case Workshop. 
\end{acknowledgments}

\bibliography{sample701}{}

@INPROCEEDINGS{Apai2019,
       author = {{Apai}, D{\'a}niel and {Milster}, Tom D. and {Kim}, Dae Wook and {Bixel}, Alex and {Schneider}, Glenn and {Rackham}, Benjamin V. and {Liang}, Rongguang and {Arenberg}, Jonathan},
        title = "{Nautilus Observatory: a space telescope array based on very large aperture ultralight diffractive optical elements}",
    booktitle = {Astronomical Optics: Design, Manufacture, and Test of Space and Ground Systems II},
         year = 2019,
       editor = {{Hull}, Tony B. and {Kim}, Dae Wook and {Hallibert}, Pascal},
       series = {Society of Photo-Optical Instrumentation Engineers (SPIE) Conference Series},
       volume = {11116},
        month = sep,
          eid = {1111608},
        pages = {1111608},
          doi = {10.1117/12.2529428},
       adsurl = {https://ui.adsabs.harvard.edu/abs/2019SPIE11116E..08A}
}

@ARTICLE{Bean2021,
       author = {{Bean}, Jacob L. and {Raymond}, Sean N. and {Owen}, James E.},
        title = "{The Nature and Origins of Sub-Neptune Size Planets}",
      journal = {Journal of Geophysical Research (Planets)},
         year = 2021,
        month = jan,
       volume = {126},
       number = {1},
          eid = {e06639},
        pages = {e06639},
          doi = {10.1029/2020JE006639},
archivePrefix = {arXiv},
       eprint = {2010.11867},
 primaryClass = {astro-ph.EP},
       adsurl = {https://ui.adsabs.harvard.edu/abs/2021JGRE..12606639B}
}

@ARTICLE{Seager2007,
       author = {{Seager}, S. and {Kuchner}, M. and {Hier-Majumder}, C.~A. and {Militzer}, B.},
        title = "{Mass-Radius Relationships for Solid Exoplanets}",
      journal = {\apj},
         year = 2007,
        month = nov,
       volume = {669},
       number = {2},
        pages = {1279-1297},
          doi = {10.1086/521346},
archivePrefix = {arXiv},
       eprint = {0707.2895},
 primaryClass = {astro-ph},
       adsurl = {https://ui.adsabs.harvard.edu/abs/2007ApJ...669.1279S}
}

@ARTICLE{Bergsten2022,
       author = {{Bergsten}, Galen J. and {Pascucci}, Ilaria and {Mulders}, Gijs D. and {Fernandes}, Rachel B. and {Koskinen}, Tommi T.},
        title = "{The Demographics of Kepler's Earths and Super-Earths into the Habitable Zone}",
      journal = {\aj},
         year = 2022,
        month = nov,
       volume = {164},
       number = {5},
          eid = {190},
        pages = {190},
          doi = {10.3847/1538-3881/ac8fea},
archivePrefix = {arXiv},
       eprint = {2209.04047},
 primaryClass = {astro-ph.EP},
       adsurl = {https://ui.adsabs.harvard.edu/abs/2022AJ....164..190B}
}

@ARTICLE{Madhusudhan2023,
       author = {{Madhusudhan}, Nikku and {Sarkar}, Subhajit and {Constantinou}, Savvas and {Holmberg}, M{\r{a}}ns and {Piette}, Anjali A.~A. and {Moses}, Julianne I.},
        title = "{Carbon-bearing Molecules in a Possible Hycean Atmosphere}",
      journal = {\apjl},
         year = 2023,
        month = oct,
       volume = {956},
       number = {1},
          eid = {L13},
        pages = {L13},
          doi = {10.3847/2041-8213/acf577},
archivePrefix = {arXiv},
       eprint = {2309.05566},
 primaryClass = {astro-ph.EP},
       adsurl = {https://ui.adsabs.harvard.edu/abs/2023ApJ...956L..13M}
}

@ARTICLE{Benneke2024,
       author = {{Benneke}, Bj{\"o}rn and {Roy}, Pierre-Alexis and {Coulombe}, Louis-Philippe and {Radica}, Michael and {Piaulet}, Caroline and {Ahrer}, Eva-Maria and {Pierrehumbert}, Raymond and {Krissansen-Totton}, Joshua and {Schlichting}, Hilke E. and {Hu}, Renyu and {Yang}, Jeehyun and {Christie}, Duncan and {Thorngren}, Daniel and {Young}, Edward D. and {Pelletier}, Stefan and {Knutson}, Heather A. and {Miguel}, Yamila and {Evans-Soma}, Thomas M. and {Dorn}, Caroline and {Gagnebin}, Anna and {Fortney}, Jonathan J. and {Komacek}, Thaddeus and {MacDonald}, Ryan and {Raul}, Eshan and {Cloutier}, Ryan and {Acuna}, Lorena and {Lafreni{\`e}re}, David and {Cadieux}, Charles and {Doyon}, Ren{\'e} and {Welbanks}, Luis and {Allart}, Romain},
        title = "{JWST Reveals CH$_4$, CO$_2$, and H$_2$O in a Metal-rich Miscible Atmosphere on a Two-Earth-Radius Exoplanet}",
      journal = {arXiv e-prints},
         year = 2024,
        month = mar,
          eid = {arXiv:2403.03325},
        pages = {arXiv:2403.03325},
          doi = {10.48550/arXiv.2403.03325},
archivePrefix = {arXiv},
       eprint = {2403.03325},
 primaryClass = {astro-ph.EP},
       adsurl = {https://ui.adsabs.harvard.edu/abs/2024arXiv240303325B}
}

@ARTICLE{Davenport2025,
       author = {{Davenport}, Brian and {Kempton}, Eliza M. -R. and {Nixon}, Matthew C. and {Ih}, Jegug and {Deming}, Drake and {Fu}, Guangwei and {May}, E.~M. and {Bean}, Jacob L. and {Gao}, Peter and {Rogers}, Leslie and {Malik}, Matej},
        title = "{TOI-421 b: A Hot Sub-Neptune with a Haze-free, Low Mean Molecular Weight Atmosphere}",
      journal = {\apjl},
         year = 2025,
        month = may,
       volume = {984},
       number = {2},
          eid = {L44},
        pages = {L44},
          doi = {10.3847/2041-8213/adcd76},
archivePrefix = {arXiv},
       eprint = {2501.01498},
 primaryClass = {astro-ph.EP},
       adsurl = {https://ui.adsabs.harvard.edu/abs/2025ApJ...984L..44D}
}

@ARTICLE{Madhusudhan2025PNAS,
       author = {{Madhusudhan}, Nikku and {Holmberg}, M{\^a}ns and {Constantinou}, Savvas and {Cooke}, Gregory J.},
        title = "{Exploring the sub-Neptune frontier with JWST}",
      journal = {Proceedings of the National Academy of Science},
         year = 2025,
        month = sep,
       volume = {122},
       number = {39},
          eid = {e2416194122},
        pages = {e2416194122},
          doi = {10.1073/pnas.2416194122},
       adsurl = {https://ui.adsabs.harvard.edu/abs/2025PNAS..12216194M}
}

@ARTICLE{Nixon2025,
       author = {{Nixon}, Matthew C. and {Somers}, R. Sander and {Savel}, Arjun B. and {Ih}, Jegug and {Kempton}, Eliza M.-R. and {Young}, Edward D. and {Schlichting}, Hilke E. and {Lichtenberg}, Tim and {Welbanks}, Luis and {Misener}, William and {Piette}, Anjali A.~A. and {Wogan}, Nicholas F.},
        title = "{Magma Ocean Interactions Can Explain JWST Observations of the Sub-Neptune TOI-270 d}",
      journal = {\apj},
         year = 2025,
        month = dec,
       volume = {995},
       number = {1},
          eid = {95},
        pages = {95},
          doi = {10.3847/1538-4357/ae17c8},
archivePrefix = {arXiv},
       eprint = {2510.07367},
 primaryClass = {astro-ph.EP},
       adsurl = {https://ui.adsabs.harvard.edu/abs/2025ApJ...995...95N}
}

@ARTICLE{Holmberg2024,
       author = {{Holmberg}, M{\r{a}}ns and {Madhusudhan}, Nikku},
        title = "{Possible Hycean conditions in the sub-Neptune TOI-270 d}",
      journal = {\aap},
         year = 2024,
        month = mar,
       volume = {683},
          eid = {L2},
        pages = {L2},
          doi = {10.1051/0004-6361/202348238},
archivePrefix = {arXiv},
       eprint = {2403.03244},
 primaryClass = {astro-ph.EP},
       adsurl = {https://ui.adsabs.harvard.edu/abs/2024A&A...683L...2H}
}

@ARTICLE{Shorttle2024,
       author = {{Shorttle}, Oliver and {Jordan}, Sean and {Nicholls}, Harrison and {Lichtenberg}, Tim and {Bower}, Dan J.},
        title = "{Distinguishing Oceans of Water from Magma on Mini-Neptune K2-18b}",
      journal = {\apjl},
         year = 2024,
        month = feb,
       volume = {962},
       number = {1},
          eid = {L8},
        pages = {L8},
          doi = {10.3847/2041-8213/ad206e},
archivePrefix = {arXiv},
       eprint = {2401.05864},
 primaryClass = {astro-ph.EP},
       adsurl = {https://ui.adsabs.harvard.edu/abs/2024ApJ...962L...8S}
}

@ARTICLE{Hu2025,
       author = {{Hu}, Renyu and {Bello-Arufe}, Aaron and {Tokadjian}, Armen and {Yang}, Jeehyun and {Damiano}, Mario and {Roy}, Pierre-Alexis and {Coulombe}, Louis-Philippe and {Madhusudhan}, Nikku and {Constantinou}, Savvas and {Benneke}, Bj{\"o}rn},
        title = "{A water-rich interior in the temperate sub-Neptune K2-18 b revealed by JWST}",
      journal = {arXiv e-prints},
         year = 2025,
        month = jul,
          eid = {arXiv:2507.12622},
        pages = {arXiv:2507.12622},
          doi = {10.48550/arXiv.2507.12622},
archivePrefix = {arXiv},
       eprint = {2507.12622},
 primaryClass = {astro-ph.EP},
       adsurl = {https://ui.adsabs.harvard.edu/abs/2025arXiv250712622H}
}

@ARTICLE{Madhusudhan2025,
       author = {{Madhusudhan}, Nikku and {Constantinou}, Savvas and {Holmberg}, M{\r{a}}ns and {Sarkar}, Subhajit and {Piette}, Anjali A.~A. and {Moses}, Julianne I.},
        title = "{New Constraints on DMS and DMDS in the Atmosphere of K2-18 b from JWST MIRI}",
      journal = {\apjl},
         year = 2025,
        month = apr,
       volume = {983},
       number = {2},
          eid = {L40},
        pages = {L40},
          doi = {10.3847/2041-8213/adc1c8},
archivePrefix = {arXiv},
       eprint = {2504.12267},
 primaryClass = {astro-ph.EP},
       adsurl = {https://ui.adsabs.harvard.edu/abs/2025ApJ...983L..40M}
}

@ARTICLE{Madhusudhan2021,
       author = {{Madhusudhan}, Nikku and {Piette}, Anjali A.~A. and {Constantinou}, Savvas},
        title = "{Habitability and Biosignatures of Hycean Worlds}",
      journal = {\apj},
         year = 2021,
        month = sep,
       volume = {918},
       number = {1},
          eid = {1},
        pages = {1},
          doi = {10.3847/1538-4357/abfd9c},
archivePrefix = {arXiv},
       eprint = {2108.10888},
 primaryClass = {astro-ph.EP},
       adsurl = {https://ui.adsabs.harvard.edu/abs/2021ApJ...918....1M}
}

@ARTICLE{Brande2024,
       author = {{Brande}, Jonathan and {Crossfield}, Ian J.~M. and {Kreidberg}, Laura and {Morley}, Caroline V. and {Barman}, Travis and {Benneke}, Bj{\"o}rn and {Christiansen}, Jessie L. and {Dragomir}, Diana and {Fortney}, Jonathan J. and {Greene}, Thomas P. and {Hardegree-Ullman}, Kevin K. and {Howard}, Andrew W. and {Knutson}, Heather A. and {Lothringer}, Joshua D. and {Mikal-Evans}, Thomas},
        title = "{Clouds and Clarity: Revisiting Atmospheric Feature Trends in Neptune-size Exoplanets}",
      journal = {\apjl},
         year = 2024,
        month = jan,
       volume = {961},
       number = {1},
          eid = {L23},
        pages = {L23},
          doi = {10.3847/2041-8213/ad1b5c},
archivePrefix = {arXiv},
       eprint = {2310.07714},
 primaryClass = {astro-ph.EP},
       adsurl = {https://ui.adsabs.harvard.edu/abs/2024ApJ...961L..23B}
}

@ARTICLE{Roy2026,
       author = {{Roy}, Pierre-Alexis and {Benneke}, Bj{\"o}rn and {Fournier-Tondreau}, Marylou and {Coulombe}, Louis-Philippe and {Piaulet-Ghorayeb}, Caroline and {Lafreni{\`e}re}, David and {Allart}, Romain and {Cowan}, Nicolas B. and {Dang}, Lisa and {Johnstone}, Doug and {Langeveld}, Adam B. and {Pelletier}, Stefan and {Radica}, Michael and {Taylor}, Jake and {Albert}, Lo{\"\i}c and {Doyon}, Ren{\'e} and {Flagg}, Laura and {Jayawardhana}, Ray and {MacDonald}, Ryan J. and {Turner}, Jake D.},
        title = "{Diversity in the haziness and chemistry of temperate sub-Neptunes}",
      journal = {Nature Astronomy},
         year = 2026,
        month = mar,
       volume = {10},
        pages = {371-384},
          doi = {10.1038/s41550-025-02723-3},
       adsurl = {https://ui.adsabs.harvard.edu/abs/2026NatAs..10..371R}
}

@ARTICLE{Fulton2017,
       author = {{Fulton}, Benjamin J. and {Petigura}, Erik A. and {Howard}, Andrew W. and {Isaacson}, Howard and {Marcy}, Geoffrey W. and {Cargile}, Phillip A. and {Hebb}, Leslie and {Weiss}, Lauren M. and {Johnson}, John Asher and {Morton}, Timothy D. and {Sinukoff}, Evan and {Crossfield}, Ian J.~M. and {Hirsch}, Lea A.},
        title = "{The California-Kepler Survey. III. A Gap in the Radius Distribution of Small Planets}",
      journal = {\aj},
         year = 2017,
        month = sep,
       volume = {154},
       number = {3},
          eid = {109},
        pages = {109},
          doi = {10.3847/1538-3881/aa80eb},
archivePrefix = {arXiv},
       eprint = {1703.10375},
 primaryClass = {astro-ph.EP},
       adsurl = {https://ui.adsabs.harvard.edu/abs/2017AJ....154..109F}
}

@ARTICLE{Kite2020,
       author = {{Kite}, Edwin S. and {Fegley}, Jr., Bruce and {Schaefer}, Laura and {Ford}, Eric B.},
        title = "{Atmosphere Origins for Exoplanet Sub-Neptunes}",
      journal = {\apj},
         year = 2020,
        month = mar,
       volume = {891},
       number = {2},
          eid = {111},
        pages = {111},
          doi = {10.3847/1538-4357/ab6ffb},
archivePrefix = {arXiv},
       eprint = {2001.09269},
 primaryClass = {astro-ph.EP},
       adsurl = {https://ui.adsabs.harvard.edu/abs/2020ApJ...891..111K}
}

@ARTICLE{Yang2013,
       author = {{Yang}, Jun and {Cowan}, Nicolas B. and {Abbot}, Dorian S.},
        title = "{Stabilizing Cloud Feedback Dramatically Expands the Habitable Zone of Tidally Locked Planets}",
      journal = {\apjl},
         year = 2013,
        month = jul,
       volume = {771},
       number = {2},
          eid = {L45},
        pages = {L45},
          doi = {10.1088/2041-8205/771/2/L45},
archivePrefix = {arXiv},
       eprint = {1307.0515},
 primaryClass = {astro-ph.EP},
       adsurl = {https://ui.adsabs.harvard.edu/abs/2013ApJ...771L..45Y}
}

@ARTICLE{Turbet2021,
       author = {{Turbet}, Martin and {Bolmont}, Emeline and {Chaverot}, Guillaume and {Ehrenreich}, David and {Leconte}, J{\'e}r{\'e}my and {Marcq}, Emmanuel},
        title = "{Day-night cloud asymmetry prevents early oceans on Venus but not on Earth}",
      journal = {\nat},
         year = 2021,
        month = oct,
       volume = {598},
       number = {7880},
        pages = {276-280},
          doi = {10.1038/s41586-021-03873-w},
archivePrefix = {arXiv},
       eprint = {2110.08801},
 primaryClass = {astro-ph.EP},
       adsurl = {https://ui.adsabs.harvard.edu/abs/2021Natur.598..276T}
}

@ARTICLE{Benneke2013,
       author = {{Benneke}, Bj{\"o}rn and {Seager}, Sara},
        title = "{How to Distinguish between Cloudy Mini-Neptunes and Water/Volatile-dominated Super-Earths}",
      journal = {\apj},
         year = "2013",
        month = {12},
       volume = {778},
       number = {2},
          eid = {153},
        pages = {153},
          doi = {10.1088/0004-637X/778/2/153},
archivePrefix = {arXiv},
       eprint = {1306.6325},
 primaryClass = {astro-ph.EP},
       adsurl = {https://ui.adsabs.harvard.edu/abs/2013ApJ...778..153B}
}

@ARTICLE{Line2016,
       author = {{Line}, Michael R. and {Parmentier}, Vivien},
        title = "{The Influence of Nonuniform Cloud Cover on Transit Transmission Spectra}",
      journal = {\apj},
         year = "2016",
        month = {3},
       volume = {820},
       number = {1},
          eid = {78},
        pages = {78},
          doi = {10.3847/0004-637X/820/1/78},
archivePrefix = {arXiv},
       eprint = {1511.09443},
 primaryClass = {astro-ph.EP},
       adsurl = {https://ui.adsabs.harvard.edu/abs/2016ApJ...820...78L}
}

@ARTICLE{Welbanks2019,
       author = {{Welbanks}, Luis and {Madhusudhan}, Nikku},
        title = "{On Degeneracies in Retrievals of Exoplanetary Transmission Spectra}",
      journal = {\aj},
         year = "2019",
        month = {5},
       volume = {157},
       number = {5},
          eid = {206},
        pages = {206},
          doi = {10.3847/1538-3881/ab14de},
archivePrefix = {arXiv},
       eprint = {1904.05356},
 primaryClass = {astro-ph.EP},
       adsurl = {https://ui.adsabs.harvard.edu/abs/2019AJ....157..206W}
}

@ARTICLE{Batalha2019,
       author = {{Batalha}, Natasha E. and {Lewis}, Taylor and {Fortney}, Jonathan J. and {Batalha}, Natalie M. and {Kempton}, Eliza and {Lewis}, Nikole K. and {Line}, Michael R.},
        title = "{The Precision of Mass Measurements Required for Robust Atmospheric Characterization of Transiting Exoplanets}",
      journal = {\apjl},
         year = 2019,
        month = nov,
       volume = {885},
       number = {1},
          eid = {L25},
        pages = {L25},
          doi = {10.3847/2041-8213/ab4909},
archivePrefix = {arXiv},
       eprint = {1910.00076},
 primaryClass = {astro-ph.EP},
       adsurl = {https://ui.adsabs.harvard.edu/abs/2019ApJ...885L..25B}
}

@ARTICLE{Welbanks2021,
       author = {{Welbanks}, Luis and {Madhusudhan}, Nikku},
        title = "{Aurora: A Generalized Retrieval Framework for Exoplanetary Transmission Spectra}",
      journal = {\apj},
         year = 2021,
        month = jun,
       volume = {913},
       number = {2},
          eid = {114},
        pages = {114},
          doi = {10.3847/1538-4357/abee94},
archivePrefix = {arXiv},
       eprint = {2103.08600},
 primaryClass = {astro-ph.EP},
       adsurl = {https://ui.adsabs.harvard.edu/abs/2021ApJ...913..114W}
}

@ARTICLE{Welbanks2024,
       author = {{Welbanks}, Luis and {Bell}, Taylor J. and {Beatty}, Thomas G. and {Line}, Michael R. and {Ohno}, Kazumasa and {Fortney}, Jonathan J. and {Schlawin}, Everett and {Greene}, Thomas P. and {Rauscher}, Emily and {McGill}, Peter and {Murphy}, Matthew and {Parmentier}, Vivien and {Tang}, Yao and {Edelman}, Isaac and {Mukherjee}, Sagnick and {Wiser}, Lindsey S. and {Lagage}, Pierre-Olivier and {Dyrek}, Achr{\`e}ne and {Triantafillides}, Anastasia},
        title = "{A high internal heat flux and large core in a warm Neptune exoplanet}",
      journal = {\nat},
         year = 2024,
        month = jun,
       volume = {630},
       number = {8018},
        pages = {836-840},
          doi = {10.1038/s41586-024-07514-w},
archivePrefix = {arXiv},
       eprint = {2405.11018},
 primaryClass = {astro-ph.EP},
       adsurl = {https://ui.adsabs.harvard.edu/abs/2024Natur.630..836W}
}

@INBOOK{Madhusudhan2018,
       author = {{Madhusudhan}, Nikku},
        title = "{Atmospheric Retrieval of Exoplanets}",
    booktitle = {Handbook of Exoplanets},
         year = 2018,
          eid = {104},
        pages = {104},
          doi = {10.1007/978-3-319-55333-7_104},
       adsurl = {https://ui.adsabs.harvard.edu/abs/2018haex.bookE.104M},
       publisher={Springer}
}

@ARTICLE{Bell2023,
       author = {{Bell}, Taylor J. and {Welbanks}, Luis and {Schlawin}, Everett and {Line}, Michael R. and {Fortney}, Jonathan J. and {Greene}, Thomas P. and {Ohno}, Kazumasa and {Parmentier}, Vivien and {Rauscher}, Emily and {Beatty}, Thomas G. and {Mukherjee}, Sagnick and {Wiser}, Lindsey S. and {Boyer}, Martha L. and {Rieke}, Marcia J. and {Stansberry}, John A.},
        title = "{Methane throughout the atmosphere of the warm exoplanet WASP-80b}",
      journal = {\nat},
         year = 2023,
        month = nov,
       volume = {623},
       number = {7988},
        pages = {709-712},
          doi = {10.1038/s41586-023-06687-0},
archivePrefix = {arXiv},
       eprint = {2309.04042},
 primaryClass = {astro-ph.EP},
       adsurl = {https://ui.adsabs.harvard.edu/abs/2023Natur.623..709B}
}

@ARTICLE{Madhusudhan2020,
       author = {{Madhusudhan}, Nikku and {Nixon}, Matthew C. and {Welbanks}, Luis and {Piette}, Anjali A.~A. and {Booth}, Richard A.},
        title = "{The Interior and Atmosphere of the Habitable-zone Exoplanet K2-18b}",
      journal = {\apjl},
         year = 2020,
        month = mar,
       volume = {891},
       number = {1},
          eid = {L7},
        pages = {L7},
          doi = {10.3847/2041-8213/ab7229},
archivePrefix = {arXiv},
       eprint = {2002.11115},
 primaryClass = {astro-ph.EP},
       adsurl = {https://ui.adsabs.harvard.edu/abs/2020ApJ...891L...7M}
}

@ARTICLE{Bixel2021,
       author = {{Bixel}, Alex and {Apai}, D{\'a}niel},
        title = "{Bioverse: A Simulation Framework to Assess the Statistical Power of Future Biosignature Surveys}",
      journal = {\aj},
         year = 2021,
        month = may,
       volume = {161},
       number = {5},
          eid = {228},
        pages = {228},
          doi = {10.3847/1538-3881/abe042},
archivePrefix = {arXiv},
       eprint = {2101.10393},
 primaryClass = {astro-ph.EP},
       adsurl = {https://ui.adsabs.harvard.edu/abs/2021AJ....161..228B}
}

@ARTICLE{Triaud2024,
       author = {{Triaud}, Amaury H.~M.~J. and {de Wit}, Julien and {Klein}, Frieder and {Turbet}, Martin and {Rackham}, Benjamin V. and {Niraula}, Prajwal and {Glidden}, Ana and {Jagoutz}, Oliver E. and {Pe{\v{c}}}, Matej and {Petkowski}, Janusz J. and {Seager}, Sara and {Selsis}, Franck},
        title = "{Atmospheric carbon depletion as a tracer of water oceans and biomass on temperate terrestrial exoplanets}",
      journal = {Nature Astronomy},
         year = 2024,
        month = jan,
       volume = {8},
       number = {1},
        pages = {17-29},
          doi = {10.1038/s41550-023-02157-9},
archivePrefix = {arXiv},
       eprint = {2310.14987},
 primaryClass = {astro-ph.EP},
       adsurl = {https://ui.adsabs.harvard.edu/abs/2024NatAs...8...17T}
}

@ARTICLE{Cherubim2025,
       author = {{Cherubim}, Collin and {Wordsworth}, Robin and {Bower}, Dan J. and {Sossi}, Paolo A. and {Adams}, Danica and {Hu}, Renyu},
        title = "{An Oxidation Gradient Straddling the Small Planet Radius Valley}",
      journal = {\apj},
         year = 2025,
        month = apr,
       volume = {983},
       number = {2},
          eid = {97},
        pages = {97},
          doi = {10.3847/1538-4357/adbca9},
archivePrefix = {arXiv},
       eprint = {2503.05055},
 primaryClass = {astro-ph.EP},
       adsurl = {https://ui.adsabs.harvard.edu/abs/2025ApJ...983...97C}
}
\bibliographystyle{aasjournalv7}
\newpage

\begin{deluxetable}{lccl}
\tabletypesize{\scriptsize}
\tablecaption{Wavelength requirements for the \nautilus{} sub-Neptune population survey. \label{scireq1}}
\tablewidth{0pt}
\tablehead{
\colhead{Requirement} & \colhead{Imaging} & \colhead{Spect.} & \colhead{Science Driver}
}
\startdata
250--350\,nm    & N/A & Optional & Stellar UV environment; atmospheric escape context \\
350--450\,nm    & N/A & Required & Rayleigh slope; cloud / aerosol diagnostics \\
450--1{,}000\,nm  & N/A & Required & Rayleigh slope; alkali features (Na, K); aerosol distinction \\
1--1.8\,$\mu$m  & N/A & Required & \ce{H2O} bands; He~I (1.083\,$\mu$m) escape diagnostic (high-$R$ option) \\
1.8--2.3\,$\mu$m & N/A & Required & \ce{H2O}, \ce{CH4} bands \\
2.3--2.9\,$\mu$m & N/A & Required & \ce{CH4}, \ce{CO} bands \\
2.9--5\,$\mu$m   & N/A & Required & \ce{CO2} (4.3\,$\mu$m), \ce{CO} (4.7\,$\mu$m); key chemistry diagnostics \\
\enddata
\tablecomments{Survey is built around transmission spectroscopy.}
\end{deluxetable}

\begin{deluxetable}{lll}
\tabletypesize{\scriptsize}
\tablecaption{System-level science requirements for the \nautilus{} sub-Neptune population survey. \label{scireq2}}
\tablewidth{0pt}
\tablehead{
\colhead{Requirement} & \colhead{Range} & \colhead{Science Driver}
}
\startdata
Photometric Filters             & Optional, multi-band            & Transit validation; host-star characterization \\
Target Brightness [mag]         & $K \lesssim 12$ (survey); $K \lesssim 9$ (benchmarks) & Per-transit SNR on warm sub-Neptunes around M/K hosts \\
Min. Spectro. Precision [ppm]   & Comparable to \jwst{} NIRISS/NIRSpec on bright sub-Neptune hosts & Resolve order-unity scale-height features \\
Image Res. [diff. limit]        &  N/A                &  \\
Min. Sky Coverage [deg$^2$]  & N/A (targeted); per-pointing FOV $\sim 30''$--$1'$  & Targeted transit hosts; FOV sized for target plus contamination reference (cf. Ariel, JWST NIRISS) \\
Min. Contrast                   & N/A                              & Transmission survey, not direct imaging \\
Spectral Resolving Power        & $R \sim 100$--$500$ (baseline); $R \gtrsim 30{,}000$ (optional) & Resolve molecular bands\\
\hline
Relevant Timescales [s]         &  transits 1--5\,h, can build to longer transit planets  &  \\
Monitoring Baseline [d]         & Months                           & Multi-transit coverage at orbital periods of days to weeks \\
Cadence [s]                     & 60--120\,s in transit            & Sample ingress/egress and transit shape \\
Rapid Response Time [s]         & Not critical (transits predictable) & --- \\
\hline
Data Volume                     & TBD                              & --- \\
Pointing Precision [arcsec]     & Sub-pixel stability per transit  & Limit systematics in transit-depth precision \\
\enddata
\tablecomments{Per-target observing strategy and total sample size scale with mission class (probe / flaglet / flagship). \emph{Uniformity} of observing strategy and reduction pipeline across the sample is itself a requirement as population inference depends on homogeneity.}
\end{deluxetable}

\end{document}